\documentclass{article}
\usepackage{spconf,amsmath,amssymb,graphicx}
\usepackage{color}
\usepackage{balance}
\usepackage{multirow}

\usepackage[]{algorithm2e}
\usepackage{wasysym}
\usepackage{bm}
\usepackage{url,cite}
\usepackage{enumitem} 
\usepackage{pifont}
\usepackage{float}
\usepackage{hyperref}

\usepackage{makecell}

\title{A study of transfer learning in music source separation}
\name{Andreas Bugler\textsuperscript{1}, Bryan Pardo\textsuperscript{1}, Prem Seetharaman\textsuperscript{1, 2}}
\address{\textsuperscript{1}Northwestern University\quad \textsuperscript{2}Descript, Inc}
\begin{document}
%
\maketitle
\begin{abstract}
Supervised deep learning methods for performing audio source separation can be very effective in domains where there is a large  amount of training data. While some music domains have enough data suitable for training a separation system, such as rock and pop genres, many musical domains do not, such as classical music, choral music, and non-Western music traditions. It is well known that transferring learning from related domains can result in a performance boost for deep learning systems, but it is not always clear how best to do pretraining. In this work we investigate the effectiveness of data augmentation during pretraining, the impact on performance as a result of pretraining and downstream datasets having similar content domains, and also explore how much of a model must be retrained on the final target task, once pretrained. 

\end{abstract}

\begin{keywords}
Audio Source Separation, Transfer Learning, Deep Clustering, Data Augmentation
\end{keywords}

\section{Introduction}
\label{sec:intro}
Music source separation is the task of isolating one or more musical sources in an audio recording of a mixture of musical sources, for example separating the vocals from accompaniment (drums, bass, and other instruments). While humans can selectively attend to the source of a musical signal in a noisy auditory scene, this task has shown itself to be much more challenging for computers to perform. Recently, progress in source separation has been driven by deep learning methods \cite{attractor, gmm_separation, dpcl, chimera, conv-tasnet}.

While deep methods are powerful, they require massive amounts of labeled, separated data, to build generalizable models. For example, U-Net for speech separation was trained on almost 2 months, or 1440 hours of audio \cite{u-net}. Music source separation suffers from a paucity of data, as music makers typically do not release their isolated tracks (stems) to the public \cite{slakh}. MUSDB \cite{musdb18}(the largest public dataset of high quality real-world recorded music) is only 15 hours long. In contrast, datasets for other computer audition tasks, such as speech-related tasks,  are much larger, on the order of 1000 hours (e.g. Librispeech \cite{librispeech}). State-of-the-art source separation systems are trained on large private datasets \cite{spleeter} but these datasets are not publicly shared, limiting who can train using that amount of data. While a large synthesized dataset does exist for rock music \cite{slakh}, it contains no vocals and the instrument sounds do not fully capture the sound qualities of real-world instruments. For many musical domains, such as classical music \cite{orchestral_separation}, choral music \cite{choir_separation}, and non-Western music traditions, data sets are orders of magnitude smaller than speech datasets.

Models that are trained on a small amount of data tend to overfit to the training set, leading to poor generalization performance. To tackle this issue of data paucity, we propose to use transfer learning for music source separation. Transfer learning is well known in the machine learning community as a way to boost performance when labeled data in the desired domain is not plentiful \cite{HowTransferrable, comprehensive_transfer}. Cramer et al. \cite{L3} found that pretraining a model on a large amount of data via audiovisual correspondence resulted in models that could be fine-tuned to downstream tasks (such as environment sound classification) with little data, resulting in  better generalization performance. 



In this work, we show that transfer learning can be an effective approach for training separation networks in limited-data regimes. We first pretrain a separation network on large available music separation datasets, such as MUSDB \cite{musdb18} and Slakh \cite{slakh}. The pretrained network is then fine-tuned to a variety of downstream separation tasks, some with as little as 5 minutes of training data. We show that these fine-tuned networks can far out-strip the performance of networks that are trained from scratch for the downstream separation task, even when the total number of training iterations is the same in both conditions. We also study the impact of data augmentation on the efficacy of transfer learning, the impact of domain mismatch between pretraining and fine-tuning tasks, and the efficacy of freezing layers of the network during fine-tuning. We will release our pretrained models and our transfer learning recipe via \emph{nussl} \cite{nussl}, an open source audio separation library.


\section{Proposed Method}
We propose to first pre-train a network to perform source separation using datasets that have ample training data. We then fine-tune the network to perform source separation on a dataset where we have very little data. We use a commonly-used network architecture - ChimeraNet \cite{chimera}. ChimeraNet is a two-headed deep architecture for source separation that simultaneously outputs deep time-frequency embeddings and time-frequency masks that can be applied to the mixture to separate sources. 


ChimeraNet is representative of both the popular mask-inference and deep clustering approaches to separation. In this work, we pretrain the deep clustering head and fine-tune both heads to the downstream separation task. While in this work, we restrict our investigation to the ChimeraNet architecture, the approach can be applied to any network architecture. Our approach consists of two stages - pre-training, and fine-tuning for transfer learning.


The goal of the separation system is to produce a mask that can be applied to the input mixture to separate a specific source. Let the audio signal be represented by it's complex Short-Time Fourier Transform (STFT), $X$. Bin $X(f, t)$ is the phase and magnitude of the the signal at time $t$ and frequency $f$. In mask inference, the task is to design a mask $M_c$, that contains real-valued scalars in $[0, 1]$, where $M_c(f, t)$ contains the proportion of the magnitude at $X(f, t)$ that comes from source $c$. 
Let $X_c = M_c \odot X$ be the STFT of the source $c$, where $\odot$ denotes element-wise multiplication.

The ChimeraNet architecture consists of a stack of recurrent layer which process the input magnitude spectrogram. The input to the network is the log-magnitude spectrogram of the mixture. There are two outputs from the network: a deep embedding space where every time-frequency point is mapped to a $D$-dimensional embedding $V$ and $N$ masks which can be applied to the mixture for separation. The embedding space is trained via the deep clustering loss:
\begin{equation}
\mathcal{L}_{\text{DC}} = ||W^{1/2}(VV^T - YY^T) W^{1/2}||^2_F.
\end{equation}
where $W$ represents a weight for every time-frequency point, $V$ are the embeddings produced by the network, and $Y$ is the ideal binary assignment of every time-frequency point.

\subsection{Pre-training}

We pre-train this network architecture using only the deep clustering loss on large datasets of music mixtures with corresponding ground truth. The mask head is left untrained. This is because while deep clustering is invariant to the number of sources that are in the mixture, mask inference is not. By training only the deep clustering head, we can adapt to different numbers of sources output by the mask inference head during fine-tuning. 

\subsection{Fine-tuning}
After pre-training, we fine-tune the network architecture on a smaller amount of task-specific data. During fine-tuning we train both the deep clustering head as well as the mask inference head. 

The mask head is trained via the mask-inference loss, using the magnitude-spectrum approximation:
\begin{equation}
\mathcal{L}_{\text{MI}} = \frac{1}{N} \Big\| |M\circ X| - |S| \Big\|_1, 
\end{equation}
where $N$ is the number of time-frequency points, $M$ is a mask, $X$ is the mixture spectrogram, and $S$ is the magnitude spectrogram of the corresponding ground truth source.

There are two options for fine-tuning: fine-tune the entire network, or only fine-tune the very last masking layer. In the first case, we apply both both the mask inference loss and the embedding loss, combining them with a sum: 
\begin{equation}
\mathcal L_{DC+MI} = \alpha \mathcal L_{DC} + (1-\alpha) \mathcal L_{MI}
\end{equation}
where $\alpha=.01$, as in prior literature \cite{chimera, bootstrapping}. We will refer to this case as \textit{whole-models}.

In the second case, we train only with $\mathcal L_{MI}$, and further only train the very last layer of the network that produces the mask. Everything preceding the mask layer in the network is frozen to the weights that were a result of pre-training. We will refer to this case as \textit{mask-models}.

\section{Experimental Design}

Our experiments are designed to investigate the impact of the proposed method for transfer learning on separation performance for a variety of downstream tasks. Downstream tasks are those for which we have little data to train a full separation system. We consider the following hypotheses: 

\textbf{H1:} We hypothesize that models that are pre-trained on a large dataset and then fine-tuned to the downstream task will have better performance than models that are trained only on the downstream task from scratch, even when the total number of training iterations is the same in both conditions.

\textbf{H2:} We hypothesize that data augmentation during pre-training will further improve the performance of transfer learning, due to increased robustness of the pre-trained model.

We also consider the following research questions:

\textbf{R1:} How does the domain of the pre-training dataset affect the performance of the fine-tuned model? If the two domains are very different, then the expectation is that fine-tuning will not result in improved performance.

\textbf{R2:} How do the two cases of fine-tuning compare in terms of performance: \textit{mask-models} versus \textit{whole-models}?

\begin{figure}[t]
    \centering
    \includegraphics[scale=.50]{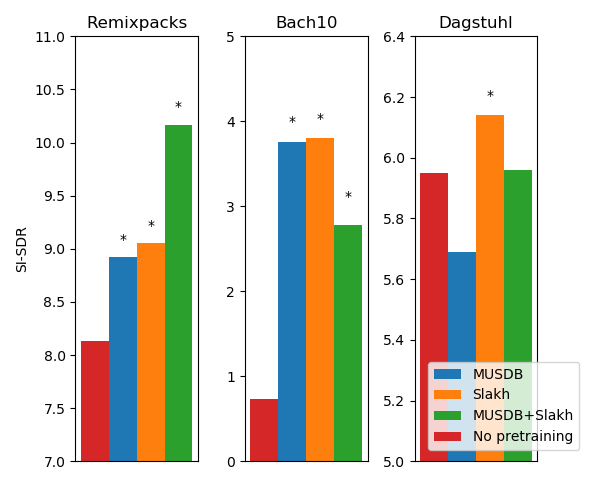}
    \vspace{-15pt}
    \caption{SI-SDR on downstream task with corresponding whole-model, with no data augmentations. We note that the smallest downstream dataset, Bach10, showed the greatest increase in separation performance. `*' over a bar represents statistically significant change over the no pretraining baseline.}
    \label{fig:si-sdrs}
    \vspace{-10pt}
\end{figure}
\begin{figure}[]
    \centering
    \includegraphics[scale=.55]{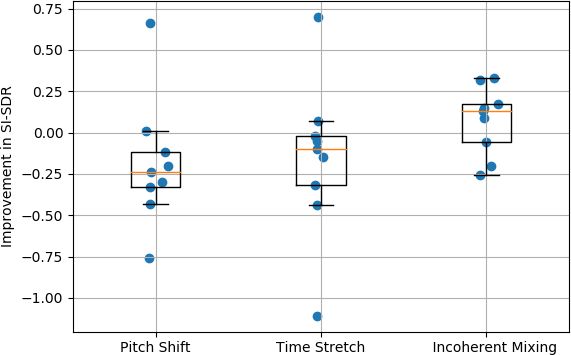}
    \vspace{-10pt}
    \caption{Improvement in SI-SDR of whole-models as a result of data augmentation during pretraining, compared to not using any data augmentation. Data augmentation does not appear to consistently result in an increase of SI-SDR, even with incoherent mixing, with median improvement of .13 SI-SDR}
    \label{fig:augmentations}
\end{figure}

\subsection{Data}
For pre-training, we use two datasets: MUSDB \cite{musdb18} and Slakh \cite{slakh}. MUSDB is a real-instrument recorded music separation dataset consisting of 15 hours of music mixtures with associated ground truth sources. There are four instruments in MUSDB: vocals, drums, bass, and all other instruments. We use the training set, which consists of 100 songs for pre-training. We group the drums, bass, and other categories as a single source: accompaniment. Slakh is a dataset consisting of synthesized instrument mixtures, generated from a large MIDI dataset. Slakh is much larger than MUSDB, with 145 hours of mixtures. We use the training set, which consists of 1890 songs. We consider pre-training on MUSDB alone, Slakh alone, and a combination of both datasets, MUSDB+Slakh. The source `other' in MUSDB+Slakh consists of the other source in MUSDB, and the piano and guitar sources in Slakh. See Table \ref{tab:dataset_info} for statistics on these datasets.

During training, we mix sources on the fly to create \emph{coherent} mixtures. Coherent mixing involves mixing stems from the same song and time frame. We can optionally use two forms of on-the-fly data augmentation during pre-training: pitch shifting and time stretching. The amount to pitch shift is chosen randomly between $-3$ and $3$ semitones. The amount of time-stretching is chosen randomly between $.8$ and $1.2$. Finally, we look at the impact of training only from incoherent mixtures. Incoherent mixtures contain stems from independently random chosen songs and time frames for each stem.

\begin{table}[t]
    \centering
    \scriptsize
\setlength\tabcolsep{1.3pt}
    \begin{tabular}{|c|c|c|c|c|c|c|}\hline
    &MUSDB&Slakh &
    \makecell{MUSDB\\+Slakh}&
    Bach10&
    \makecell{Dagstuhl\\ChoirSet} &  Remixpacks\\\hline
         Training &600 &7830&8430&  5 & 70 & 192\\\hline
         Testing & 300&870&1170&.5& 7 & 96\\\hline
         Sources & \makecell{Accomp. \\ Vocals}&
         \makecell{Piano+Guitar\\ Drums}
         &\makecell{Bass\\Drums\\Other}
         &\makecell{Violin\\ Saxophone\\ Bassoon\\ Clarinet}& 
         \makecell{Male\\ Female} & \makecell{Snare\\ Kick Drum \\ Cymbals}\\\hline
    \end{tabular}
    \caption{Duration in minutes of each dataset's training and testing splits in minutes and dataset sources.}
    \vspace{-10pt}
    \label{tab:dataset_info}
\end{table}

We consider three datasets for downstream tasks. These datasets are all small, and are generally not suitable for deep learning due to how little data there is. The three datasets are Bach10 \cite{bach10}, Dagstuhl ChoirSet \cite{dagstuhl}, and Remixpacks. See Table \ref{tab:dataset_info} for duration and sources in each dataset. In Dagstuhl, the male source is generated from summing the bass and tenor sources, and the female source is similarly generated from summing the alto and soprano parts. Only songs where all four parts are existent and recorded with dynamic mics are used in training and evaluation in Dagstuhl. Remixpacks is a dataset containing stems scraped from \href{https://remixpacks.ru}{https://remixpacks.ru}. The stems are incoherently mixed in Remixpacks during training and testing, while stems in Bach10 are mixed incoherently during training only. Remixpacks is mixed incoherently because finding a song with perfectly aligned tracks was rare, and Bach10 was incoherently mixed during training for better performance.

\subsection{Training procedure and evaluation}

We use a ChimeraNet model with a BLSTM stack with $500$ hidden units, $4$ layers, and $D=20$. The STFT is calculated with a hop length of $128$, window length of $512$, and the square root of the Hann window as the window function, which results in $257$ frequency bins in the STFT. All of our experiments are done at a sampling rate of 16kHz. All models are trained with AutoClip \cite{autoclip}, with $p=10$, a learning rate of $.001$ with the Adam optimizer. The learning rate is halved if the validation loss averaged across every $100$ iterations plateaus after $500$ iterations.

To investigate \textbf{H1}, we first pre-train 3 networks: one on MUSDB, one on Slakh, and one on MUSDB+Slakh. Each pre-trained network is fine-tuned to each downstream dataset, resulting in a total of 9 fine-tuned models. In addition to these three networks, we train a baseline model on each downstream dataset alone, from scratch. This results in $12$ models. The pre-trained models are all trained for $10000$ iterations, with a batch size of $24$, on $10$-second mixtures. For fine-tuning, the models are trained for an additional $2000$ iterations on the data for the final (downstream) task. The baseline model is trained for $12000$ iterations, only on the data for the final task. Finally, for each downstream dataset we fine-tune two variants: the entire network (whole-model) or only the mask (mask-model).

To investigate \textbf{H2}, we pre-train with and without data augmentation under four conditions: no data augmentation, applying time stretching only, applying pitch shifting only, and applying incoherent mixing only.

To evaluate separation performance, we use the scale-invariant source-to-distortion ratio (SI-SDR) \cite{half-baked}. Higher values indicate better separation performance. Each model we train is evaluated on $1000$ samples of test set corresponding to the downstream dataset trained on. We measure the average SI-SDR over these $1000$ samples. We use a one-sided Wilcoxon Signed Rank Test to calculate statistical significance. To reject the null hypothesis, we set a threshold of $p<.001$.

\begin{table}[t]\centering
\scriptsize
\label{tab:mask}
\begin{tabular}{lr|rrrr}
& &Remixpacks &Dagstuhl &Bach10 \\\hline
\multirow{3}{*}{No Augmentations} &MUSDB &-4.04 &-1.27 &-5.05 \\		
&Slakh &-2.32 &-1.76 &-3.43 \\		
&MUSDB+Slakh &-4.78 &-3.03 &-4.94 \\\hline
\multirow{3}{*}{Pitch Shift} &MUSDB &-5.34 &\textbf{1.16} &-3.78 \\		
&Slakh &-2.80 &-2.29 &-4.72 \\		
&MUSDB+Slakh &-2.38 &-2.02 &-4.91 \\\hline	
\multirow{3}{*}{Time Stretch} &MUSDB &-4.08 &\textbf{0.85} &-4.68 \\		
&Slakh &-3.33 &-0.34 &-3.60 \\		
&MUSDB+Slakh &-1.69 &-2.17 &-3.99 \\\hline
\multirow{3}{*}{Incoherent Mixing} &MUSDB &-4.50 &\textbf{0.69} &-4.57 \\		
&Slakh &-3.40 &-0.10 &-2.32 \\		
&MUSDB+Slakh &-1.75 &-0.63 &-4.85 \\		

\end{tabular}
\caption{Difference in SI-SDR between the mask-model and the corresponding whole-model. Bolded values indicate statistically significant increase in SI-SDR.}
\end{table}

\section {Results}
Figure \ref{fig:si-sdrs} shows the effect of starting from pre-trained models for Bach10, Remixpacks, and Dagstuhl. For both Bach10 and Remixpacks, we find that the fine-tuned models far our-perform models that start from scratch. For Dagstuhl, the effect is much less, with only the model that is pre-trained from Slakh out-performing the baseline. This may be due to domain mismatch between the dataset used for pre-training and the downstream task. Dagstuhl consists of all vocals sources, whereas the MUSDB data, for example, groups all vocals into one source. This domain mismatch may be the cause of reduced performance on this downstream task.

\begin{figure}[]
\vspace{-10pt}
    \centering
    \includegraphics[scale=.5]{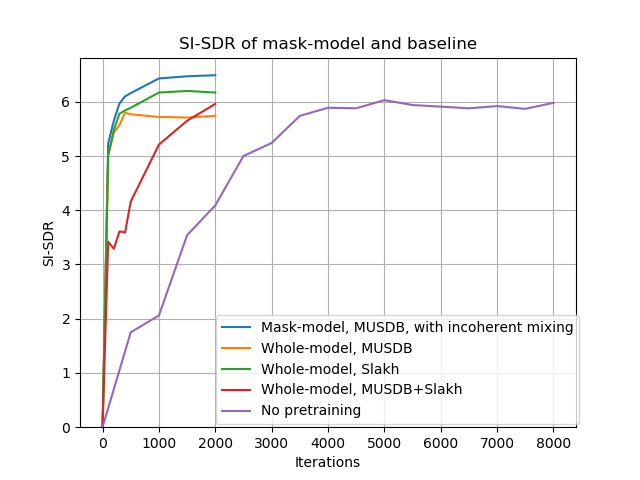}
    \vspace{-10pt}
    \caption{On Dagstuhl ChoirSet: test SI-SDR at each training step. Starting from a pretrained model reaches higher performance and gets there faster than from scratch.}
    \label{fig:iterations}
\end{figure}

Figure \ref{fig:iterations} shows the performance of each model every 100 iterations, however, that fine-tuning still reaches equivalent or slightly better performance much faster than training from scratch. The best model that was fine-tuned to Dagstuhl took only $100$ iterations to reach ~5dB SI-SDR, whereas the baseline took $2500$ iterations to reach that same performance. This indicates that starting from a pre-trained model not only matches or exceeds the performance of models trained from scratch, but also that it is more efficient in terms of iterations.

In Figure \ref{fig:augmentations}, we show the effect of applying data augmentations during pre-training. We observe that the impact of data augmentation during training is ambiguous at best, with very little change when comparing against the baseline model. The best augmentation was incoherent mixing, which gave a $.13$ SI-SDR boost. Finally, in Table \ref{tab:mask}, we can see that freezing all but the mask layer during fine-tuning results in a drop in performance, except for the specific case where a Dagstuhl fine-tuned model is pretrained on MUSDB, and a data augmentation is applied.

\section{Conclusion}
In this work, we investigated whether transfer learning can be effective for training source separation models in regimes where data is highly limited. We proposed a specific transfer learning recipe, where a model is pre-trained on a large dataset using a deep clustering objective function, and then fine-tuned to a small dataset. We show that this is effective, out-performing baselines that were trained from scratch on small datasets, enabling the training of source separation models that work on instruments for which there is not much data. In future work, we plan to explore other forms of pre-training, including self-supervised learning, transferring models from other audio domains (e.g. speech), and expanding the set of down-stream datasets we build models for. We will release our pre-trained models and fine-tuning recipe via nussl \cite{nussl}, allowing both researchers to experiment with transfer learning for source separation, and practitioners to easily apply source separation to their own fields.

\bibliographystyle{IEEEbib}
\bibliography{bibliography}
\end{document}